\documentclass[10pt]{article}
\usepackage{latexsym}
\usepackage[dvips]{graphicx,color}
\newcommand{\be}{\begin{equation}}
\newcommand{\ee}{\end{equation}}
\def\n{\noindent}
\catcode `\@=11
\catcode `\@=12
\begin{document}
\begin{center}
\large{\bf { A New Class of LRS Bianchi Type-I Cosmological        
       Models in Lyra's Manifold}} \\
\vspace{10mm}
\normalsize{\bf{Anirudh Pradhan $^1$, Dinkar Singh Chauhan$^2$}} \\
\normalsize{$^1$Department of Mathematics, Hindu Post-graduate College, 
Zamania-232 331, Ghazipur, India.} \\
\normalsize{E-mail : pradhan@iucaa.ernet.in, acpradhan@yahoo.com}\\
\normalsize{$^2$Department of Mathematics, Prithwi Raj Chauhan Degree College,
Bhiti-275 101, Mau, U. P., India.} \\
\end{center}
\vspace{10mm}
\begin{abstract} 
LRS Bianchi type-I models have been studied in the cosmological
theory based on Lyra's geometry. A new class of exact solutions has
been obtained by considering a time dependent displacement field for 
variable deceleration parameter models of the universe. We have compared 
our models with those of Einstein's field theory with the cosmological 
term $\Lambda$. Our frame of reference is restricted to the recent Ia 
observations of supernovae. Some physical behaviour of the models is 
also examined in the presence of perfect fluids.
\end{abstract}
\smallskip
\n Key Words : Cosmology; L R S Bianchi type-I Models; Lyra Geometry\\
\n PACS number : 98.80.-k, 98.80.Jk

\section{Introduction}

   In 1917 Einstein introduced the cosmological constant into his field
equations in order to obtain a static cosmological model since, as is
well known, without the cosmological term his field equations admit only
non-static solutions. After the discovery of the redshift of galaxies
and its explanation as being due to the expansion of the universe,
Einstein regretted his introduction of the cosmological constant.
Recently, there has been much interest in the cosmological term
in context of quantum field theories, quantum gravity, super-gravity
theories, Kaluza-Klein theories and the inflationary-universe 
scenario. Shortly after Einstein's general theory of relativity Weyl, 
in 1918, suggested the first so-called unified  field theory based on a 
generalization of Riemannian geometry. In retrospect, it would seem more
appropriate to call Weyl's theory a geometrized theory of gravitation and 
electromagnetism (just as the general theory was a geometrized theory of
gravitation only), rather than a unified field theory. It is not quite clear
to what extent the two fields have been unified, even though they acquire 
(different) geometrical significances in the same geometry. The theory was 
never taken seriously because it was based on the concept of non integrability
of length transfer, and, as pointed out by Einstein, this implies that 
spectral frequencies of atoms depend on their past histories and therefore 
have no absolute significance. Nevertheless, Weyl's geometry provides an 
interesting example of non-Riemannian connections, and recently 
Folland \cite{ref1} has given a global formulation of Weyl manifolds thereby 
clarifying considerably many of Weyl's basic ideas.     

   In 1951 Lyra \cite{ref2} proposed a modification of Riemannian geometry
by introducing a gauge function into the structureless manifold,
as a result of which the cosmological constant arises naturally from
the geometry. This bears a remarkable resemblance to Weyl's geometry. But in
Lyra's geometry, unlike Weyl's, the connection is metric preserving as in
Remannian; in other words, length transfers are integrable. Lyra also 
introduced the notion of a gauge and in the ``normal'' gauge the curvature 
scalar in identical to that of Weyl. In consecutive investigations 
Sen \cite{ref3}, Sen and Dunn \cite{ref4} proposed a new scalar-tensor 
theory of gravitation and constructed an analog of the Einstein field 
equations based on Lyra's geometry. It is thus possible \cite{ref3} to 
construct a geometrized theory of gravitation and electromagnetism much 
along the lines of Weyl's ``unified'' field theory without, however, the 
inconvenience of non-integrability length transfer.\\

   Halford \cite{ref5} has pointed out that the constant vector displacement
field $\phi_i$ in Lyra's geometry plays the role of cosmological
constant $\Lambda$ in the normal general relativistic treatment. It
is shown by Halford \cite{ref6} that the scalar-tensor treatment based on
Lyra's geometry predicts the same effects, within observational limits,
as the Einstein's theory. Several authors Sen and Vanstone \cite{ref7}, 
Bhamra \cite{ref8}, Karade and Borikar \cite{ref9}, Kalyanshetti and 
Wagmode \cite{ref10}, Reddy and Innaiah \cite{ref11}, Beesham \cite{ref12}, 
Reddy and Venkateswarlu \cite{ref13}, Soleng \cite{ref14}, have studied 
cosmological models based on Lyra's manifold with a constant displacement 
field vector. However, this restriction of the displacement field to be 
constant is merely one of convenience and there is no {\it a priori} reason 
for it. Beesham \cite{ref15} considered FRW models with time dependent 
displacement field. He has shown that by assuming the energy density of 
the universe to be equal to its critical value, the models have the $k=-1$ 
geometry. Singh and Singh \cite{ref16}, Singh and Desikan \cite{ref17} have 
studied Bianchi-type I, III, Kantowaski-Sachs
and a new class of cosmological models with time dependent displacement
field and have made a comparative study of Robertson-Walker models
with constant deceleration parameter in Einstein's theory with
cosmological term and in the cosmological theory based on Lyra's geometry.
Soleng \cite{ref14} has pointed out that the cosmologies based on Lyra's 
manifold with constant gauge vector $\phi$ will either include a creation  
field and be equal to Hoyle's creation field cosmology \cite{ref18} or 
contain a special vacuum field which together with the gauge 
vector term may be considered as a cosmological term. In the latter case the 
solutions are equal to the general relativistic cosmologies with a 
cosmological term. 

    The Einstein's field equations are a coupled system of highly nonlinear
differential equations and we seek physical solutions to the field equations
for their applications in cosmology and astrophysics. In order to solve the 
field equations we normally assume a form for the matter content or that 
space-time admits killing vector symmetries \cite{ref19}. Solutions to the field 
equations may also be generated by applying a law of variation for Hubble's 
parameter which was proposed by Berman \cite{ref20}. In simplest case the 
Hubble law yields a constant value for the deceleration parameter. It is 
worth observing that most of the well-known models of Einstein's theory and 
Brans-Deke theory with curvature parameter $k=0$, including inflationary models, 
are models with constant deceleration parameter. In earlier
literature cosmological models with a constant deceleration
parameter have been studied by several authors \cite{ref21}.
But redshift magnitude test has had a chequered history. During the 1960s 
and the 1970s, it was used to draw very categorical conclusions. The 
deceleration parameter $q_{0}$ was then claimed to lie between $0$ and $1$ 
and thus it was claimed that the universe is decelerating. Today's situation,
we feel, is hardly different. Observations (Knop et al. \cite{ref22};
Riess et al., \cite{ref23}) of Type Ia Supernovae (SNe) allow to probe the
expansion history of the universe. The main conclusion of these
observations is that the expansion of the universe is accelerating. So we 
can consider the cosmological models with variable cosmological term and 
deceleration parameter. The readers are advised to see the papers by 
Vishwakarma and Narlikar \cite{ref24} and Virey et al. \cite{ref25} and 
references therein for a review on the determination of the deceleration 
parameter from Supernovae data.

Recently Pradhan and Otarod \cite{ref26} have studied the universe 
with time dependent deceleration parameter and $\Lambda$-term in 
presence of perfect fluid. Motivated with the situation discussed above, 
in this paper, we study a new class of LRS Bianchi-I cosmological models 
in Lyra geometry by considering a time dependent deceleration parameter 
in an expanding universe. This paper is organized as follows. The metric and 
the field equations are presented in Section 2. In Section 3 we deal with a 
general solution. The Sections 4, 5, and 6 contain the three different cases 
for the solutions in exponential, polynomial and sinusoidal forms respectively. 
Finally in Section 7 concluding remarks will be given.  
\section{The Metric and Field Equations}
We consider LRS Bianchi type I space-time
\begin{equation}
\label{eq1}
ds^2 = dt^2 -  A^2dx^2 - B^2 (dy^2 + dz^2)
\end{equation}
where $A = A (x,t), B=B(x,t)$.
We take a perfect fluid form for the energy momentum tensor
\label{eq2}
\begin{equation}
T_{ij} = (p+\rho) u_i u_j - p g_{ij}
\end{equation}
together with co-moving coordinates $u^i u_i =1$, where $u_i = 
(0, 0, 0, 1)$.\\ 
The field equations in normal gauge for 
Lyra's manifold, as obtained by Sen [3] are 
\label{eq3}
\begin{equation}
R_{ij} - \frac{1}{2} g_{ij} R + \frac{3}{2} \phi_i \phi_j
- \frac{3}{4} g_{ij} \phi_k \phi^k = - 8 \pi G T_{ij} 
\end{equation}
where $\phi$ is a time-like displacement field vector defined by
$\phi_i = (0,0,0,\beta(t))$ and other symbols have their
usual meaning as in Riemannian geometry. Here we want to mention the fact
that the ansatz choosing the coordinate system with matter require the
vector field happens to be in the required form exactly in the matter 
comoving coordinates. The essential difference between the cosmological
theories based on Lyra geometry and the Riemannian geometry lies in the fact
that constant vector displacement field $\beta$ arises naturally from the 
concept of gauge in Lyra geometry where as cosmological constant $\Lambda$ 
was introduced in {\it adhoc} fashion in the usual treatment. 

Now the field equations can be set up and one obtains
\begin{equation}
\label{eq4}
 \frac{2\ddot B}{B} + \frac{\dot B^2}{B^2} - \frac{B'^2}{A^2 B^2}
+ \frac{3}{4} \beta^2 = - \chi p
\end{equation}
\begin{equation}
\label{eq5}
\dot B' - \frac{B' \dot A}{A} = 0 
\end{equation}
\begin{equation}
\label{eq6}
\frac{\ddot A}{A} + \frac{\ddot B}{B} + \frac{\dot A \dot B}
{AB} - \frac{B''}{A^2 B} + \frac{A' B'}{A^3 B}
+ \frac{3}{4} \beta^2 = -\chi p
\end{equation}
\begin{equation}
\label{eq7}
\frac{2B''}{A^2 B} - \frac{2 A' B'} {A^3 B} + \frac{B'^2}
{A^2 B^2} - \frac{2 \dot A \dot B}{AB} - \frac{\dot B^2}{B^2}
+ \frac{3}{4} \beta^2 = \chi\rho.
\end{equation}
The energy conservation equation is
\begin{equation}
\label{eq8}
\chi \dot\rho + \frac{3}{2} \beta \dot\beta + \left[\chi (p + 
\rho)+ \frac{3}{2} \beta^2\right] \left(\frac{\dot A}{A} + 
\frac{2 \dot B}{B}\right) = 0
\end{equation}
where $\chi = 8 \pi G$. Here and in what follows a prime and a 
dot indicate partial differentiation with respect to $x$ and $t$
respectively. Equations (\ref{eq4})-(\ref{eq7}) are four equations 
in five unknowns $A, B, \beta, p$ and $\rho$. For complete determinacy 
of the system one extra condition is needed. One way is to use an 
equation of state. The other alternative is a mathematical assumption 
on the space-time and then to discuss the physical nature of the universe. 
In this paper we confine ourselves to assume an equation of state
\begin{equation}
\label{eq9}
p = \gamma \rho, ~ 0 \leq \gamma \leq 1
\end{equation}
\section{Solutions of the field equations}
Equation (\ref{eq5}), after integration, yields
\begin{equation}
\label{eq10}
A = \frac{B'}{\l}
\end{equation}
where $\l$ is an arbitrary function of $x$. Equations (\ref{eq4}) and 
(\ref{eq6}), with the use of equation (\ref{eq10}), reduces to
\begin{equation}
\label{eq11}
\frac{B}{B'}\left(\frac{\ddot B}{B}\right)' + \frac{\dot B}{B'}
\frac{d}{dt} \left(\frac{B'}{B}\right) + \frac{\l^2}{B^2}
\left(1 - \frac{B \l'}{B'\l}\right) = 0
\end{equation}
Setting
\begin{equation}
\label{eq12}
\frac{B'}{B} = \mbox{functions of $x$}
\end{equation}
Equation (\ref{eq11}) yields on  integration
\begin{equation}
\label{eq13}
B = \l S(t),
\end{equation}
where $S(t)$ is an arbitrary function of $t$. With the help of equation 
(\ref{eq13}), equation (\ref{eq10}) becomes
\begin{equation}
\label{eq14}
A = \frac{\l'}{\l} S
\end{equation}
Now the metric (\ref{eq1}) takes the form
\begin{equation}
\label{eq15}
ds^2 = dt^2 - S^2(t) [dX^2 + e^{2X} (dy^2 + dz^2)],
\end{equation}
where $X = \ln \l$. Equations (\ref{eq4}) and (\ref{eq4}) give
\begin{equation}
\label{eq16}
\chi p = \frac{1}{S^2} - 2\frac{\ddot S}{S} - \frac{\dot S^2}{S^2}
- \frac{3}{4} \beta^2
\end{equation}
\begin{equation}
\label{eq17}
\chi\rho = \frac{3 \dot S^2}{S^2} - \frac{3}{S^2} - \frac{3}{4} \beta^2
\end{equation}
Using equation (\ref{eq4}) and eliminating $\rho(t)$ from equations (\ref{eq16}) 
and (\ref{eq17}) we have
\begin{equation}
\label{eq18}
\frac{2 \ddot S}{S} +(1+3 \gamma) \frac{\dot S^2}{S^2} - (1+3 \gamma)
\frac{1}{S^2} + \frac{3}{4} (1-\gamma) \beta^2 = 0
\end{equation}
Now the expressions for the energy density and the pressure are given
by
\begin{equation}
\label{eq19}
\chi p = \chi \gamma \rho = \frac{4 \gamma}{(1-\gamma)} \left[\frac{\dot S^2}
{S^2} - \frac{1}{S^2} + \frac{\ddot S}{2S}\right]
\end{equation}
The function $S(t)$ remains undetermined. To obtain its explicit 
dependence on $t$, one may have to introduce additional assumptions.
Accordingly we assume the deceleration parameter to be variable and set
\begin{equation}
\label{eq20}
q = - \frac{S \ddot S}{{\dot S}^2} = - \left(\frac{\dot H + H^2}{H^2}\right) 
= b \mbox{(variable)},
\end{equation}
where $H = \dot S/S$ is the Hubble parameter. The above equation may be 
rewritten as
\begin{equation}
\label{eq21} \frac{\ddot{S}}{S}+ b \frac{{\dot{S}}^2}{S^{2}} = 0.
\end{equation}
The general solution of Eq. (\ref{eq21}) is given by
\begin{equation}
\label{eq22} \int{e^{\int{\frac{b}{S}dS}}}dS = t + k,
\end{equation}
where $k$ is an integrating constant.

In order to solve the problem completely, we have to choose
$\int{\frac{b}{S}dS}$ in such a manner that Eq. (\ref{eq22})
be integrable.

Without loss of generality, we consider
\begin{equation}
\label{eq23} \int{\frac{b}{S}dS} = {\rm ln} ~ {L(S)},
\end{equation}
which does not effect the nature of generality of solution. Hence
from Eqs. (\ref{eq22}) and (\ref{eq23}), we obtain
\begin{equation}
\label{eq24} \int{L(S)dS} = t + k.
\end{equation}
Of course the choice of $L(S)$, in Eq. (\ref{eq24}), is quite
arbitrary but, since we are looking for physically viable models
of the universe consistent with observations, we consider the
following case:
\section{Solution in the Exponential Form}
Let us consider $ L(S) = \frac{1}{k_{1}S}$, where $k_{1}$ is arbitrary constant.

In this case on integration of  Eq. (\ref{eq24}) gives the exact solution 
\begin{equation}
\label{eq25} S(t) = k_{2}e^{k_{1}t}, 
\end{equation}
where $k_{2}$ is an arbitrary constant. Using Eqs. (\ref{eq9}) and (\ref{eq25}) 
in (\ref{eq18}) and (\ref{eq19}), we obtain expressions for displacement field 
$\beta$, pressure $p$ and energy density $\rho$ as 
\begin{equation}
\label{eq26} \beta^{2} = \frac{4(1 + 3\gamma)}{3(1 - \gamma)k_{2}^{2}e^{2k_{1}t}} 
- \frac{4(1 + \gamma)k_{1}^{2}}{(1 - \gamma)},  
\end{equation} 
\begin{equation}
\label{eq27} \chi p = \chi \gamma \rho = \frac{4}{(1 - \gamma)}\left[\frac{3}{2}
k_{1}^{2} - \frac{1}{k_{2}^{2}e^{2k_{1}t}}\right].
\end{equation}
\begin{figure}[htbp]
\centering
\includegraphics[width=8cm,height=8cm,angle=0]{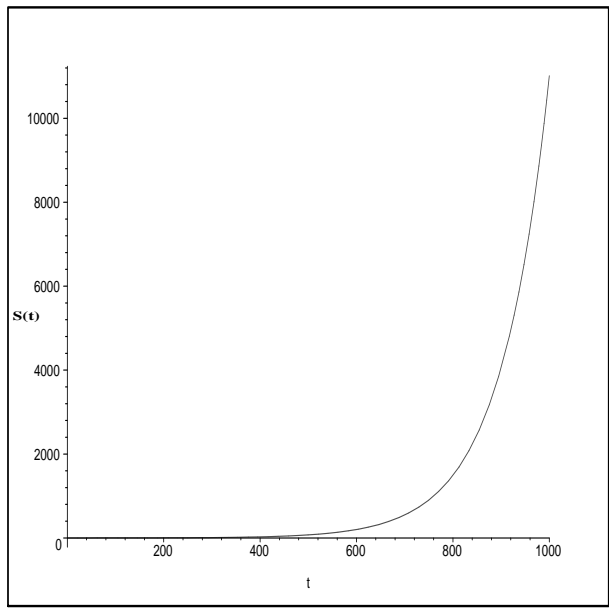}
\caption{The plot of scale factor $S(t)$ vs time with parameters 
$k_1 = 0.01$, $k_2 = 0.5$, and $\gamma = 0.5$}
\end{figure}
From Eq. (\ref{eq25}), since scale factor can not be negative, we find $S(t)$ 
is positive if $k_{2} > 0$. From Figure 1, it can be seen that 
in the early stages of the universe, i.e., near $t = 0$, the scale factor of the 
universe had been approximately constant and had increased very slowly. At  
specific time the universe had exploded suddenly and expanded to large 
scale. This is consistent with Big Bang scenario.

From Eq. (\ref{eq26}), it is observed that $\beta^{2}$ is a decreasing function of 
time. As mentioned earlier the constant vector displacement field $\phi_i $ in Lyra's 
geometry plays the role of cosmological constant $\Lambda$ in the normal general 
relativistic treatment and the scalar-tensor treatment based on Lyra's geometry 
predicts the same effects, within observational limits, as the Einstein's theory. 
Recent cosmological observations (Garnavich et al. \cite{ref27}, Perlmutter et al. 
\cite{ref28}, Riess et al. \cite{ref29}, Schmidt et al. \cite{ref30}) suggest the 
existence of a positive cosmological constant $\Lambda$ with the magnitude 
$\Lambda(G\hbar/c^{3}\approx 10^{-123}$. These observations on magnitude and 
red-shift of type Ia supernova suggest that our universe may be an accelerating 
one with induced cosmological density through the 
cosmological $\Lambda$-term. In our model, it is seen that $\beta$ plays 
the same role as cosmological constant and preserves the same character as 
$\Lambda$-term, in turn with recent observations.

The expressions for $\beta^{2}$ and $\rho$ cannot be determined for the empty 
universe $(p = \rho = 0)$ and stiff matter $(p = \rho)$ models. In this case, 
the Ricci scalar is obtained as 
\begin{equation}
\label{eq28} R = \frac{6(k_{1}^{2}k_{2}e^{k_{1}t} - 1)}{k_{2}^{2}e^{2k_{1}}} 
+ 6 k_{1}^{2}.   
\end{equation}
From above equation, we see that the Ricci scalar remains positive for
$$k_{1} >\frac{1}{\sqrt{k_{2}(1 + k_{2})}}.$$  
The expansion and shear scalar are given by
\begin{equation}
\label{eq29} \theta = 3k_{1}, ~ ~ ~ \sigma = 0.   
\end{equation}
The model represents uniform expansion as can be seen from Eq. (\ref{eq29}). 
The flow of the fluid is geodetic with the acceleration vector 
$f_{i} = (0, 0, 0, 0)$.
\section{Solution in the Polynomial Form}
Let $ L(S) = \frac{1}{2k_{3}\sqrt{S + k_{4}}}$, where $k_{3}$ and $k_{4}$ 
are  constants. \\ 
In this case, on integrating, Eq. (\ref{eq24}) gives the exact solution
\begin{equation}
\label{eq30}
S(t) = \alpha_{1}t^{2} + \alpha_{2}t + \alpha_{3},
\end{equation} 
where $\alpha_{1}$, $\alpha_{2}$ and $\alpha_{3}$ are arbitrary constants. 
Using Eqs. (\ref{eq9}) and (\ref{eq30}) in (\ref{eq18}) and (\ref{eq19}), we 
obtain the expressions for displacement field $\beta$, pressure $p$ and 
energy density $\rho$ as 
\begin{equation}
\label{eq31} \beta^{2} = - \frac{4(1 + 3\gamma)[8\alpha_{1}(\alpha_{1}t + 
\alpha_{2})t + 4\alpha_{1}\alpha_{3} + \alpha_{2}^{2} -1]}{(1 - \gamma)
(\alpha_{1}t^{2} + \alpha_{2}t + \alpha_{3})^{2}},  
\end{equation} 
\begin{equation}
\label{eq32} \chi p = \chi \gamma \rho =  \frac{4[5\alpha_{1}(\alpha_{1}t 
+ \alpha_{2})t + \alpha_{1}\alpha_{3} + \alpha_{2}^{2} -1]}{(1 - \gamma)
(\alpha_{1}t^{2} + \alpha_{2}t + \alpha_{3})^{2}}.  
\end{equation}
\begin{figure}[htbp]
\centering
\includegraphics[width=8cm,height=8cm,angle=0]{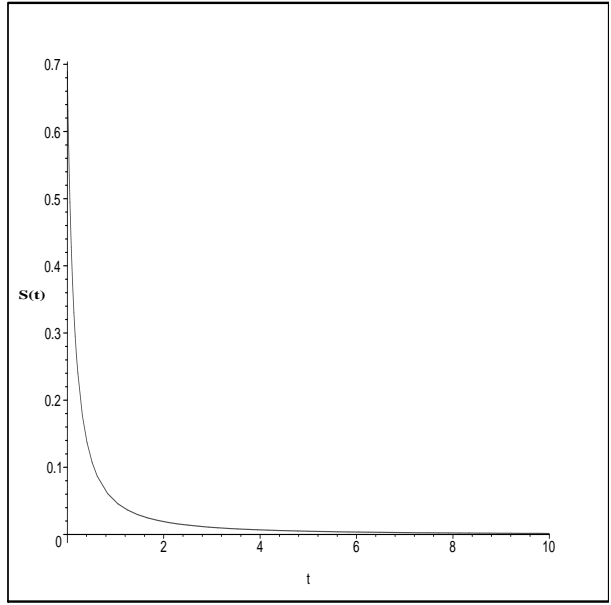}
\caption{The plot of scale factor $S(t)$ vs time with parameters 
$\alpha_1 = 1.00$, $\alpha_2 = 4.00$, $\alpha_3 = 1.00$ and $\gamma = 0.5$}
\end{figure}

From Eq. (\ref{eq30}), we note that $S(t) > 0$ for $0 \leq t < \infty$ 
if $\alpha_{1}$, $\alpha_{2}$ and $\alpha_{3}$ are positive constants.  
Figure $2$, shows that the scale factor is a decreasing function of 
time, implying that our universe is expanding.

Eq. (\ref{eq31}) shows that $\beta^{2} < 0$ for all times as 
$\gamma - 1 < 0$ and is a decreasing function of time, characteristically 
similar to $\Lambda$ in Einstein's theory of gravitation. In this model, 
$\beta$ plays the role as cosmological constant and it preserves the same 
character as $\Lambda$-term. This is consistent with recent observations 
(Garnavich et al. \cite{ref27}, Perlmutter et al. \cite{ref28}, Riess et al. 
\cite{ref29}, Schmidt et al. \cite{ref30}). A negative cosmological constant 
adds to the attractive gravity of matter; therefore, universe with a negative 
cosmological constant is invariably doomed to re-collapse. A positive 
cosmological constant resists the attractive gravity of matter due to 
its negative pressure. For most of the time, the positive cosmological constant 
eventually dominates over the attraction of matter and drives the universe to 
expand exponentially.

The expressions for $\beta^{2}$ and $\rho$ cannot be 
determined for the empty universe $(p = \rho = 0)$ and stiff matter 
$(p = \rho)$ models. In this case, the Ricci scalar is obtained as 
\begin{equation}
\label{eq33}
R = \frac{6[(2\alpha_{1}t + \alpha_{2})^{2} + 2\alpha_{1} - 1]}
{(\alpha_{1}t^{2} + \alpha_{2}t + \alpha_{3})^{2}}.
\end{equation}
From Eq. (\ref{eq33}) we observe that Ricci scalar remains positive 
if $\alpha_{2}^{2} > \frac{1 - 2\alpha_{1}}{6}$. This condition also implies 
that $\alpha_{1} < \frac{1}{2}$.
The scalar of expansion is given by
\begin{equation}
\label{eq34} \theta = \frac{3(2\alpha_{1} t + \alpha_{2})}{(\alpha_{1}t^{2} 
+ \alpha_{2}t + \alpha_{3})}.   
\end{equation}
It is of the interest to note  that all physical parameters in our models are 
defined at $t = 0$ and we do not have any singularity.


\section{Solution in the Sinusoidal Form}
Let $ L(S) = \frac{1}{\beta\sqrt{1 - S^{2}}}$, where $\beta$ is constant. \\
In this case, on integrating, Eq. (\ref{eq24}) gives the exact solution
\begin{equation}
\label{eq35}
S = M\sin(\beta t) + N\cos(\beta t) + \beta_{1},
\end{equation} 
where $M$, $N$ and $\beta_{1}$ are constants.
Using Eqs. (\ref{eq9}) and (\ref{eq35}) in (\ref{eq18}) and (\ref{eq19}), we 
obtain the expressions for displacement field $\beta$, pressure $p$ and 
energy density $\rho$ as 
\begin{equation}
\label{eq36} \beta^{2} = \frac{4\left[(1 + 3\gamma)\{1 -\beta^{2}(M\cos{\beta t} - 
N\sin{\beta t})^{2}\} + 2\beta^{2}Q(Q- \beta_{1})\right]}{3(1 - \gamma)Q^{2}},  
\end{equation} 
\begin{equation}
\label{eq37} \chi p = \chi \gamma \rho = \frac{2\left[\beta^{2}\{P - 6MN\sin{\beta t}
\cos{\beta t} - \beta_{1}(Q - \beta_{1})\} - 2\right]}{(1 - \gamma)Q^{2}},    
\end{equation} 
where 
$$Q = M\sin{\beta t} + N \cos{\beta t} + \beta_{1},$$
$$ P = (2M^{2} - N^{2})\cos{\beta t}^{2} + (2N^{2} - M^{2})\sin{\beta t}^{2}.$$
\begin{figure}[htbp]
\centering
\includegraphics[width=8cm,height=8cm,angle=0]{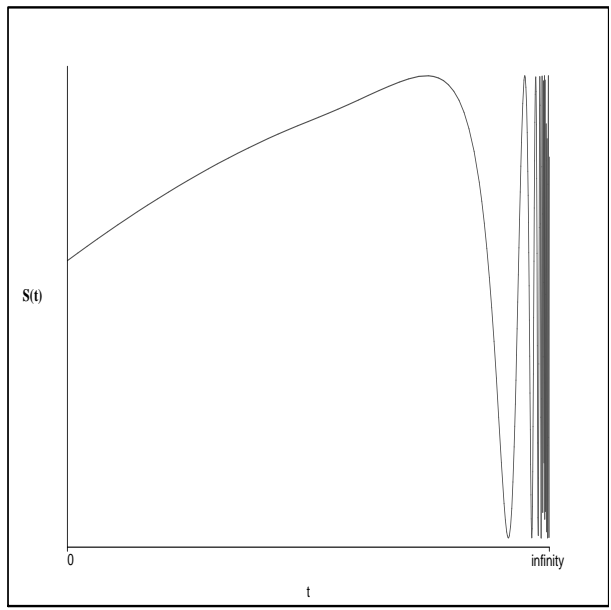}
\caption{The plot of scale factor $S(t)$ vs time with parameters 
$M = 2.00$, $N = 1.00$, $\beta = 10.00$, $\beta_1 = 0.2$, and $\gamma = 0.5$}
\end{figure}
In this case the Ricci scalar and the scalar of expansion are obtained as
\begin{equation}
\label{eq38}
R = \frac{6\left[\left(M\beta \cos(\beta t) - N\beta \sin(\beta t)\right)^{2}
 - \beta^{2}\left(M\sin(\beta t) + N\cos(\beta t)\right) - 1\right]}
{\left(M\sin(\beta t) + N\cos(\beta t) + \beta_{1}\right)^{2}},
\end{equation}
\begin{equation}
\label{eq39} \theta = \frac{3\beta[M\cos{\beta t} - N\sin{\beta t}]}
{M\sin{\beta t} + N\cos{beta t} + \beta-{1}}.
\end{equation}
From the Figure 3, we note that at early stage of the universe, the scale 
of the universe increases gently and then decreases sharply, and after wards it 
will oscillate for ever. We must mention here that the oscillation takes 
place in positive quadrant this has physical meaning. Since, in this case, we 
have many alternatives for choosing values of $M$, $N$, 
$\beta$, $\beta_{1}$, it is just enough to look for suitable values of these 
parameters, such that the physical initial and boundary conditions are satisfied. 
\section{Conclusions} 
In this paper we have obtained exact solutions of Sen equations in Lyra 
geometry for time dependent deceleration parameter. The nature of the 
displacement field $\beta(t)$ and the energy density $\rho(t)$ have been 
examined for three cases (i) exponential form (ii) polynomial form and 
(iii) sinusoidal form . The solutions obtained in Sections $4$, $5$ and $6$
are to our knowledge quite new. Here the displacement field $\beta$ plays 
the role of a variable cosmological term $\Lambda$.
  
	 Recently there is an upsurge of interest in scalar fields in 
general relativity and alternative theories of gravitation in the context 
of inflationary cosmology. Therefore the study of cosmological models
in Lyra geometry are relevant for inflationary models. Further the space
dependence of the displacement field $\beta$ is important for inhomogeneous
models in the early stage of the evolution of the universe. Besides, the 
implication of Lyra's geometry for bodies of  astrophysical interest  is 
still an open question. The problem of equations of motion and gravitational 
radiation need investigation.
\section*{Acknowledgement}
One of the authors (A. Pradhan) would like to thank the Inter-University Centre 
for Astronomy and Astrophysics, Pune, India for providing hospitality where part 
of this work was carried out.

\end{document}